\documentclass[showpacs,twocolumn,floatfix,superscriptaddress,prb,eqsecnum]{revtex4}
\usepackage{graphicx,subfigure}
\usepackage{amsmath}

\def\tr{\mathop{\rm tr}}

\def\be#1\ee{\begin{equation}#1\end{equation}}
\newcommand{\ba}{\begin{eqnarray} }
\newcommand{\ea}{\end{eqnarray} }

\def\b{\boldsymbol b}

\begin{document}
\title{Maximal positive cross shot noise from Andreev reflection}
\author{A. Bednorz}
\affiliation{Institute of Theoretical Physics, University of Warsaw, ul.~Ho\.{z}a 69, PL 00-681 Warszawa, Poland}
\affiliation{Fachbereich Physik, Universit{\"a}t Konstanz, D-78457 Konstanz, Germany}
\author{J. Tworzyd{\l}o}
\affiliation{Institute of Theoretical Physics, University of Warsaw, ul.~Ho\.{z}a 69, PL 00-681 Warszawa, Poland}
\author{J. Wr{\'o}bel}
\affiliation{Institute of Physics, Polish Academy of Sciences, al.~Lotnik\'ow 32/46, PL 02-668 Warszawa, Poland}
\author{T. Dietl}
\affiliation{Institute of Theoretical Physics, University of Warsaw, ul.~Ho\.{z}a 69, PL 00-681 Warszawa, Poland}
\affiliation{Institute of Physics, Polish Academy of Sciences, al.~Lotnik\'ow 32/46, PL 02-668 Warszawa, Poland}

\date{\today}

\begin{abstract}
The current flowing from a superconductor to a two-terminal setup describing a nanostructure
connected to normal-metal leads is studied. 
We provide an example of scattering matrix giving ideal splitting off
electrons from a Cooper pair by means of Cauchy-Bunyakovsky-Schwarz inequality.  The proposal of the junction and its possible variants are discussed in a context of possible experiments.

\end{abstract}

\maketitle
\section{Introduction}

For more than a decade it has been accepted that the
phenomena of quantum transport in mesoscopic systems
are intimately connected with the fermionic nature of carriers,
both electrons and holes.
The simplest fingerprint of statistical interaction is the
sub-poissonian shot noise measured in the quantum point contact.\cite{butt} The electronic Hanbury Brown-Twiss experiment\cite{HanTwi}
performed in the quantum Hall regime\cite{butt2,neg}
demonstrates strong anti-correlation of current fluctuations,
the effect known as anti-bunching,
characteristic for fermionic particles obeying the Pauli exclusion principle.


Quantum transport in a hybrid normal metal--superconductor (NS) junction involves yet another quasiparticle: the Cooper pair carrying an effective charge $2e$. In the Andreev reflection regime\cite{andat}
$eV \ll \Delta$ (where $V$ is the applied bias, $\Delta$ is
the superconducting gap)
the electron incident from the normal part of the junction
is reflected as a hole. The remaining charge $2e$ is absorbed
into the superconductor region as a single Cooper pair.
The effective charge of Cooper pairs leads to doubling
of NS junction conductance\cite{butt} and also doubles
the Fano factor of a tunneling barrier at the NS interface.\cite{bee02}


It has been proposed few years ago\cite{tor00} that the statistical
properties of Cooper pairs can be investigated
in the hybrid NS junction in a way analogous to the Hanbury Brown-Twiss
experiment. Fermionic anti-bunching still gives negative \emph{particle} cross correlations
but the charge reversal in Andreev reflection may lead to positive \emph{charge} or current correlations. The possibility to detect positive cross-correlation
in the NS junction\cite{tor00,marti,schech,faz,bign,duh}
is by no means a trivial prospect, as charge Andreev reflection is limited to energies within the superconducting gap. Still, the cartoon picture one keeps
in mind is that of two Cooper-pair partners undergoing a separation into different leads and then subject to a correlation measurement.



The cross-correlations are limited by Cauchy-Bunyakovsky-Schwarz (CBS) inequality
which states that cross correlations never exceed autocorrelation.
We shall call ideal splitting the situation when CBS inequality is saturated.
Theoretical analysis of positive cross-correlations in NS junctions have addressed two simple geometries so far. In particular, in the presence of many modes in the leads chaotic mode mixing dominates, so that
random matrix theory (RMT) can be employed.\cite{Sam00}. The magnitude of positive correlations
is much smaller than  CBS limit in this regime. The
experimental realization reported in Ref.~\onlinecite{Cha01} is believed to
be in the RMT regime, and so the effect remains elusive.
Another geometry is the original proposal of the Y-shaped
junction supporting only a single mode in the normal leads.
Theoretical predictions for this geometry do not give ideal splitting.\cite{Mar00}
In the recent paper,
\cite{bee2008}
it has been shown that the CBS limit can be indeed realized
at the edge of topological insulator. The limit can be interpreted as an ideal
splitting 
off electrons from a Cooper pair.

We look for a general condition for the CBS limit, in particular in the case of
single-mode terminal. We also point out that correlations can be positive also
at finite temperature without voltage bias.
An example of ideal splitting 
is attainable in a simple setup -- $X$-junction, without any bound states \cite{bee2008} or external 
filters.\cite{martles,bay,bou,san}
The junction has two branches coupled to the superconductor, where
all the branches support only a single mode.
At first sight it might seem that this geometry consists of
just two $Y$-junctions. However,
by tuning parameters (width or length) in the middle part
of the $X$ junction, it is possible to find a
Ramsauer-Townsend resonance.\cite{mott} We show that
the cross-correlations are maximized
in the superconducting case and vanish
in the normal case, provided the junction is
tuned at the resonance. 



With the progress in gating and with the advent of technology resulting in smooth
interfaces,\cite{Wro06} nanostructures containing a small and controllable number of 
modes, which is required in the $X$-junction should become available in the near future. The positive cross-correlations 
being a coherent quantum effect are expected to be more pronounced
in these devices.

%



We begin by defining ideal splitting off a Cooper pair -- maximal cross
correlations. Next, we look for the condition of ideal splitting in the
case of two normal terminals. The case of zero temperature limit is
highlighted. We provide general expressions for the current and noise for
single mode terminals. Then the $X$-junction is presented as a realization
of ideal splitting. Finally we show numerical results for transport in the
ideal junction and discuss possible modifications.

\section{Ideal splitting}

The average current and  zero-frequency correlation function in a many-terminal junction can be
written as long time averages of transferred charge,
\ba
&&\bar{I}_i=\langle I_i(t)\rangle=\sum_{\alpha\in i}q_\alpha\langle N_\alpha\rangle/t_0,\label{avn}\\
&&S_{ij}=2\int dt\langle \delta I_i(t)\delta I_j(0)\rangle=\mathop{\sum_{\alpha\in i}}_{\beta\in j}
2q_\alpha q_\beta\langle \delta N_\alpha\delta N_\beta\rangle/t_0,\nonumber
\ea
for $\delta I=I-\bar{I}$. Here $N_\alpha$ denotes the number of particles
with the charge $q_\alpha$ transferred to the mode $\alpha$ in time $t_0$.
The summations are performed over all modes and particle types at the given terminal ($i$ or $j$).
The dependence of the long time particle transfer  on the scattering matrix
can be derived using full counting statistics,\cite{les,naz} generalized to the case of
NS interface, \cite{muz,bel2} presented in detail in Appendix A.

In both cases, normal and superconducting, the noise magnitude satisfies the CBS inequality
(\ref{Schwarzmaster}), which yields
\be
S^2_{12}\leq S_{11}S_{22}\label{schwa}.
\ee
Additionally, at zero temperature and equal bias voltage at $1$ and $2$,
we have the total noise magnitude $2S_0=S_{11}+S_{12}+S_{21}+S_{22}$ satisfying
\be
2S_0(k_BT=0)\leq\left\{\begin{array}{rr}
|e(\bar{I}_1+\bar{I}_2)|&\mbox{ in normal case}\\
2|e(\bar{I}_1+\bar{I}_2)|&\mbox{ in superconducting case}
\end{array}\right.\label{subpf}
\ee
which follows from (\ref{subp}), as we show in Appendix A. 

We shall call \emph{ideal splitting} the case when $S_{12}=S_{11}=S_{22}$.
Additionally, one can maximize $S_{12}/(\bar{I}_1+\bar{I}_2)$ which is limited
by $e$ at zero temperature.

\section{Scattering matrix of an NS junction}

\begin{figure}
\includegraphics[scale=.6]{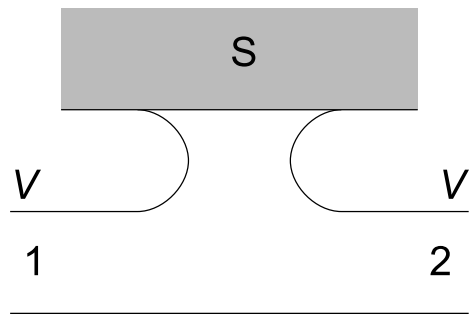}
\caption{The NS junction connected to the superconductor (S) and two normal terminals $1$ and $2$.}\label{supcon}
\end{figure}

The use standard scattering formalism for dynamics of charged quasiparticles. \cite{butt}
Fermionic operators for incoming
and outgoing states, $\psi^{in}$ and $\psi^{out}$, respectively, are
decomposed into modes, $\psi=\sum_n \psi_nc_n$, where $\psi_n$ is the normalized wavefunction of the mode and $c_n$ is the mode annihilation operator. The modes
are related by unitary scattering matrix $s$, with $c_n^{out}=\sum_ms_{nm}c_m^{in}$.

We shall consider a junction between superconductor (ground) and two normal terminals 
at the same voltage $V$ in the way presented in Fig.~\ref{supcon}.
The reference point for energies is the middle of the superconducting gap.
We include holes and Andreev reflection that converts electrons into holes and vice versa.\cite{and}.

Electrons and holes move independently far from the superconductor. Accordingly, if the electron wavefunction is $\psi$,  the corresponding one for holes is $\psi^\ast$.
Hence, in the normal region, the scattering matrix for electrons, $s_0$ determines the one for holes, $s_0^\ast$.

The superconductor mixes, however, electrons and holes.
The resulting scattering matrix can be written in the form
\be
s=\left(\begin{array}{ll}
s_{ee}&s_{eh}\\
s_{he}&s_{hh}\end{array}\right)\;.\label{seh}
\ee
In the normal region (no transitions between holes and electrons),
\be
s_N=\left(\begin{array}{ll}
s_0&0\\
0&s_0^\ast\end{array}\right)\;.
\ee
On the other hand, the pure effect of superconductor -- Andreev reflection, ignoring dynamics in the normal part, is
\be
s_A=-i\left(\begin{array}{ll}
0&e^{i\phi}\\
e^{-i\phi}&0\end{array}\right)\;,
\ee
where $\phi$ is the macroscopic phase of the superconductor,
and we assume energies $E\ll \Delta$, where $2\Delta$ is the superconducting gap and  is counted
for the middle of the gap.

We decompose
normal scattering matrix
\be
s_0=\left(\begin{array}{ll}
r&t\\
t^T&r'\end{array}\right),
\ee
where the submatrix $r$ describes reflection for normal leads, $r'$ -- for the leads connected to the
superconductor while $t$ and $t^T$ -- transmission between normal and superconducting leads.
Ignoring energy dependence and taking into account unitarity of $s_0$, we obtain\cite{butt,bee02,bee}
\ba
&&s_{ee}=(1+rr^\dag)^{-1}(r+r^T+rr^\dag(r-r^T))\nonumber\\
&&s_{eh}=-i e^{i\phi}(1+rr^\dag)^{-1}(1-rr^\dag)\nonumber\\
&&s_{he}=-i e^{-i\phi}(1+r^\dag r)^{-1}(1-r^\dag r)\label{supm}\\
&&s_{hh}=(1+r^\dag r)^{-1}(r^\ast+r^\dag+r^\dag r(r^\ast-r^\dag))\;.\nonumber
\ea
Hence, the whole transport properties are described by $r$.
We show in Appendix B that the ideal splitting is possible only at zero temperature.

\section{Single mode terminals}
Looking for ideal splitting,
we restrict ourselves to the case of symmetric single mode terminals.
so that
$r$ depends only on two complex parameters,
\be
r=\left(\begin{array}{ll}
A&B\\
B&A\end{array}\right)\;. \label{rmm}
\ee
In this case at zero temperature the ideal splitting yields the condition $A=0$,
as we show in Appendix B.
A similar matrix has been used in the previous calculations.\cite{tor00}
However, the authors could not get ideal splitting since it is not possible in a three-mode
$Y$ junction (see Appendix B).

For $A=0$, the matrix in the superconducting case has the simple form
\ba
&&s_{ee}=s_{hh}^\ast=\frac{2B}{1+|B|^2}\left(\begin{array}{rr}0&1\\1&0\end{array}\right),\nonumber\\
&&s_{eh}=-s_{he}^\ast
=-ie^{i\phi}\frac{1-|B|^2}{1+|B|^2}I.\label{scmat}
\ea
The form of the scattering matrix $s$ given by (\ref{seh}) and (\ref{scmat}) describes
a process of either transmitting an electron to the neighboring terminal or
reflecting it as a hole. We will use this simple interpretation later.
In this case, using (\ref{avn}) and (\ref{nne}), we have

\ba
&&\bar{I}_1=\frac{4e^2V}{h}(1-|B|^2)^2/(1+|B|^2)^2,\\
&&S_{11}=S_{12}=\frac{32e^2|eV|}{h}
|B|^2(1-|B|^2)^2/(1+|B|^2)^4. \nonumber
\ea
The magnitude of cross correlation is positive and maximal with respect to the CBS inequality in this case. In the limit $|B|\to 1$
we have $S_{12}\to 2|e\bar{I}_1|$. The maximum  value of $S_{12}$ is $2e^2|eV|/h$ for $|B|^2=3-\sqrt{8}$.
Importantly, cross noise vanishes in the normal case.

\begin{figure}
\subfigure{
\includegraphics[scale=.5]{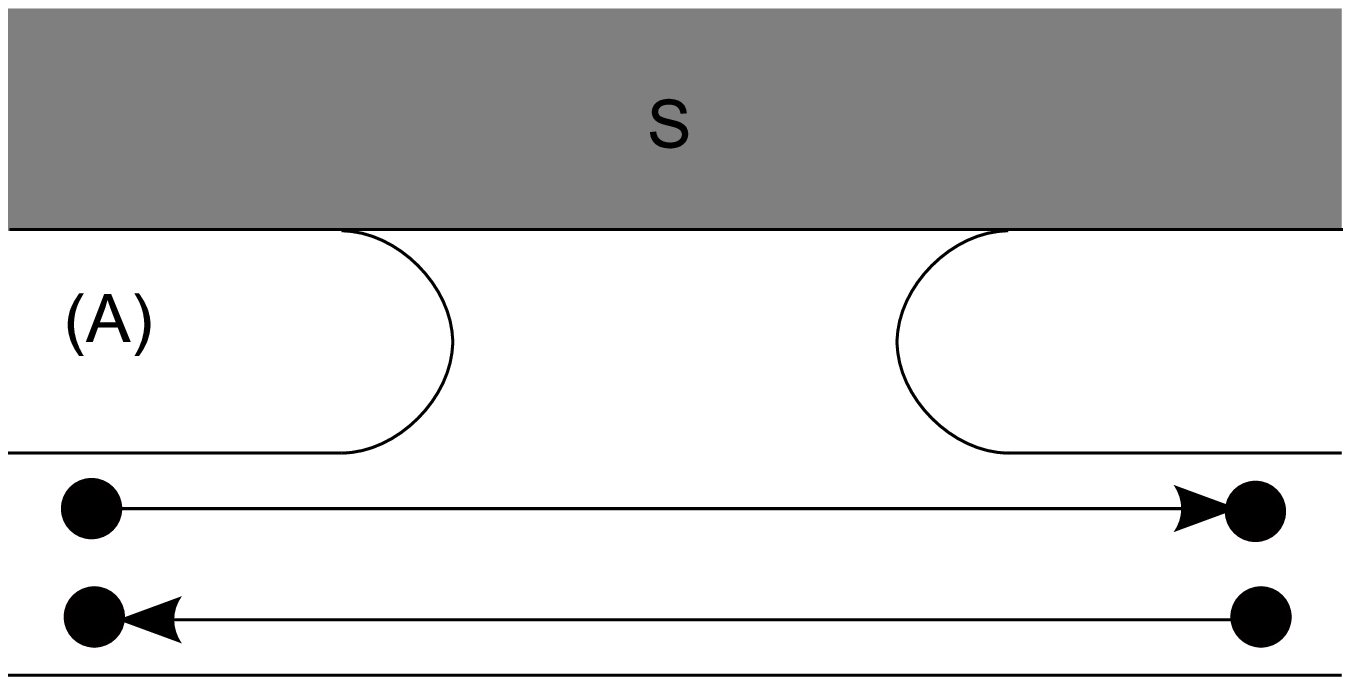} }
\subfigure{
\includegraphics[scale=.5]{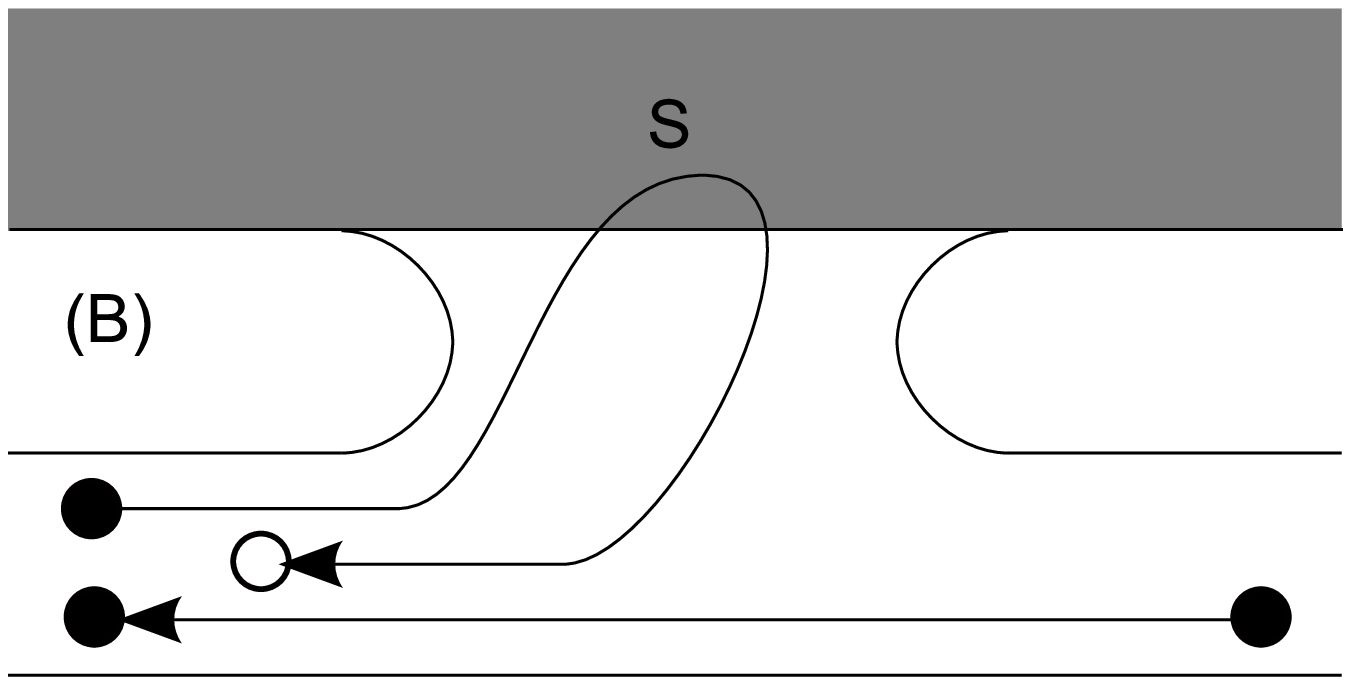}}
\subfigure{
\includegraphics[scale=.5]{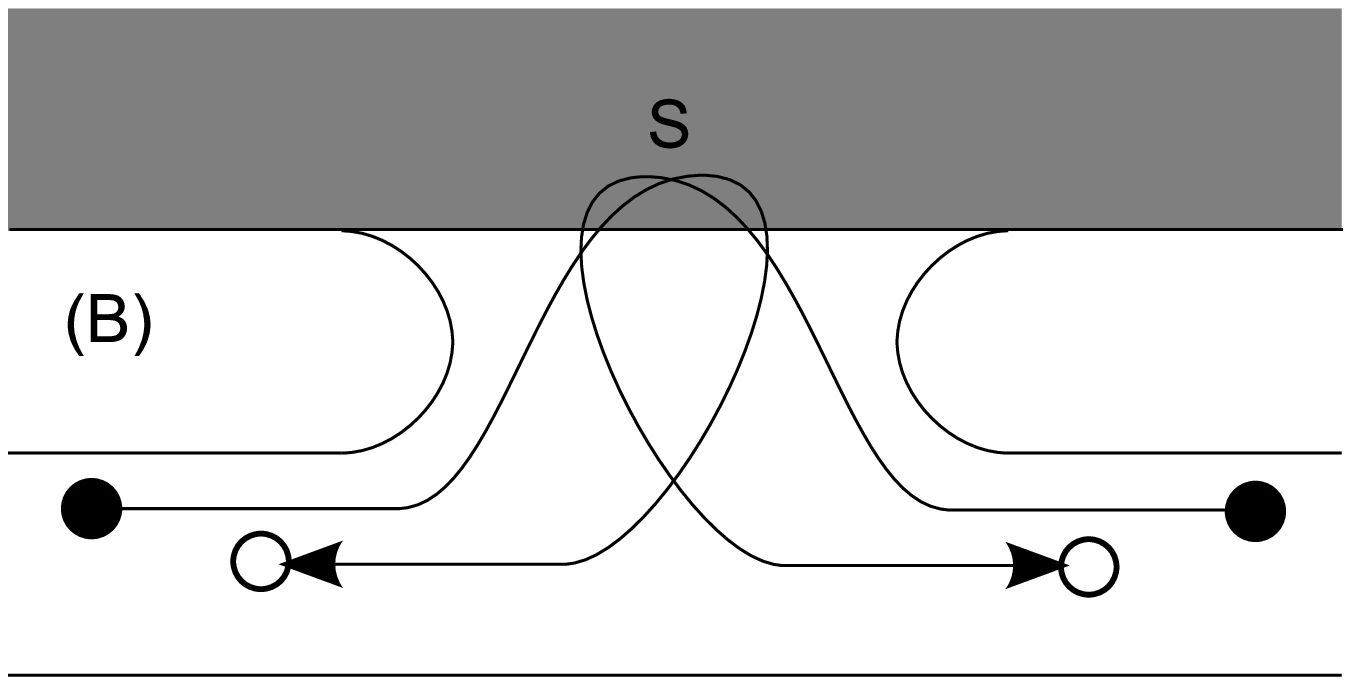}}
\caption{The three transport events at ideal splitting. Electrons are black and holes are white.(A) Both electrons pass. (B) One electron is converted into a hole.
(C) Both electrons are converted.}\label{tre}
\end{figure}

A transparent interpretation of our results can be given in terms of full counting statistics.
We can use the event counting (\ref{prob}) for the scattering matrix (\ref{scmat}), keeping in mind that incoming charge has to be subtracted while outgoing -- added.
Each electron incoming to the terminal $1$ can be either sent to the terminal $2$ or backscattered
to $1$ as a hole. For electrons incoming to $2$, one has only to exchange the role of terminals.
So, for each pair of electrons at $1$ and $2$, there are three possibilities, depicted in Fig.~\ref{tre}: (A) Both electrons are sent to
the neighbor terminal with zero charge flow. (B) One of electrons is converted into a hole with charges $+e$
going out of the junction at each side. (C) Both electrons are converted into holes with charges $+2e$ at each side.
The cross correlations are certainly positive.

In general, there could be other examples of ideal splitting, involving 
many-mode terminals. In principle, starting from the condition (\ref{cond}) one
can find the constraints for $r$. However, the procedure becomes lengthy for large
matrices.

\section{Average and shot noise}
We shall derive the general formulas for the current and noise in the case
of the junction described by reflection submatrix (\ref{rmm}).
We begin with the junction with all normal terminals, $1,2$ at voltage $V$, while the superconductor is replaced by ground. We assume that the scattering matrix is constant for energies in $[E_F,E_F+eV]$ and $eV\ll E_F$. We consider only electrons (without holes). For each mode, there are two spin orientations.
We shall use parameters $A$ and $B$ defined  in (\ref{rmm}), $C=|A|^2+|B|^2$, $D=1-C$, $w(x)=x\mathrm{cth}(x)$, $v=eV/k_BT$.
Using Eqs.~(\ref{nne}) and (\ref{avn}) we obtain
\ba
&&\bar{I}_1=\bar{I}_2=2e^2VD/h,\:S_{12}=S_{21}=-\frac{8e^2k_BT}{h}(|B|^2+\nonumber\\
&&4(w(v/2)-1)\mathrm{Re}^2 AB^\ast),\\
&&S_{11}=S_{22}=\frac{8e^2k_BT}{h}(D^2+|B|^2+w(v/2)CD).
\nonumber
\ea

In the case of NS junction two spin orientations will be replaced
by two particle types - electrons and holes. The bias voltage $V$ is counted in reference to the middle of the superconductor gap, and it has the opposite effect on electrons and holes.
We assume $|eV|,k_BT\ll \Delta$.
We shall express mean current and noise by elements of matrices $r$
and $R=rr^\ast$.
In our special case of the symmetric junction, we have $r=r^T$ and matrices $r$,$r^\ast$  and $R$ commute.
From Eqs.~(\ref{supm}) and  (\ref{nne}) we have
\ba
&&\bar{I}_1=\bar{I}_2=\frac{4e^2V}{h}\left(\frac{(1-R)^2}{(1+R)^2}\right)_{11},\nonumber\\
&&S_{11}=S_{22}=\frac{8k_BT}{h}(a+b(w(v)-1)),\label{noi}\\
&&S_{12}=S_{21}=\frac{8k_BT}{h}(c+d(w(v)-1)).\nonumber
\ea
Here
\ba
&&a=1-4\left|\left(\frac{r}{1+R}\right)_{11}\right|^2+\left[\left(\frac{1-R}{1+R}\right)_{11}\right]^2,\nonumber\\
&&b=\left(\frac{4R}{(1+R)^2}\right)_{11}\left(\frac{(1-R)^2}{(1+R)^2}\right)_{11}
\!\!\!\!+4\left|\left(\frac{r(1-R)}{(1+R)^2}\right)_{11}\right|^2\!\!,\nonumber\\
&&c=\left[\left(\frac{1-R}{1+R}\right)_{12}\right]^2-4\left|\left(\frac{r}{1+R}\right)_{12}\right|^2,\\
&&d=4\left|\left(\frac{r(1-R)}{(1+R)^2}\right)_{12}\right|^2-16\left[\left(\frac{R}{(1+R)^2}\right)_{12}\right]^2.\nonumber
\ea

Denoting conductance by $G=\bar{I}_{1}/V$, the total noise has the form
\be
S_0=4k_BTG(1+F(w(\tilde{v})-1)),
\ee
with $\tilde{v}=v/2$ and $\tilde{v}=v$ in the normal and superconducting case, respectively.
The conductance $G$ and Fano factor $F\in[0,1]$ are different in both cases.

The values of cross correlation $S_{12}$ are always negative in the normal case, but they can be either negative or positive
in the superconducting case. Interestingly, $S_{12}$ will be positive even at $V=0$  for positive $c$. This happens
at $A=\exp(i\phi)(3-\sqrt{3})/2$ and $B=\exp(i\phi)(1-\sqrt{3})/2$ giving $c=+1/8$. This is possible for a $Y$ junction.
However, this possibility is not mentioned in Ref.~\onlinecite{tor00} as only zero temperature case
is there considered.

\section{Resonance in the $X$-junction}

Now, we would like to find a realistic geometry leading to the reflection submatrix (\ref{rmm}) with $A=0$. We need
at least two modes to be later connected to the superconductor, which is realized by the $X$-junction presented below.

Let us consider the following problem:  how to find a potential that for a four-terminal
junction and suitable geometry gives the scattering matrix without backscattering, {\em i.~e.} with zero reflection amplitudes?
This special case of scattering is often (mostly in three dimensions) referred to as
Ramsauer-Townsend (RT) resonance.\cite{mott}

\begin{figure}
\includegraphics[scale=.5]{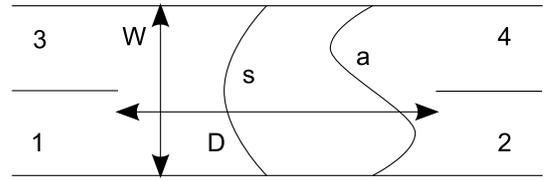}
\caption{$X$-wire geometry with four modes: symmetric (s) and antisymmetric (a) modes
are depicted. The upper, lower, and middle horizontal lines mark the infinite potential which
pins down the wavefunction.}\label{fii}
\end{figure}

We shall present such an example, starting from the usual two-dimensional
Schr\"odinger equation
\be
E_F\psi=-\frac{\hbar^2}{2m}\Delta\psi+V\psi
\ee
and the potential
\be
V=\left\{\begin{array}{ll}
+\infty &\mbox{ for }y\notin[0,W]\mbox{ or }y=W/2,x\notin[0,D]\\
0&\mbox{ otherwise }\end{array}\right.
\ee
for the junction presented in Fig.~\ref{fii}.
The symmetry helps to reduce the number of parameters describing the junction.
For the considered values of the Fermi energy $E_F=\hbar^2k_F^2/2m$  only single modes
in the terminals and only two modes in the middle part of the junction are occupied,  $k_F\in]2\pi/W,3\pi/W[$.
It is convenient to introduce the  wavenumbers $k_1=\sqrt{k_F^2-\pi^2/W^2}$ and $k_2=\sqrt{k_F^2-4\pi^2/W^2}$,
which are real and positive in the considered range of $k_F$ values. Far from the junction the wavefunction has the form
\be
\psi(x,y)=\sum_j (a_j\psi_j^{in}(x,y)+b_j\psi_j^{out}(x,y)).
\ee
Here $\psi_j^{in}=e^{i\epsilon_j k_1x}|\sin(2\pi y/ W)|\theta_j(y)$ with $\epsilon_{1,3}=+1$,
$\epsilon_{2,4}=-1$, $\theta_{1,2}(y)=\theta(y)\theta(W/2-y)$, $\theta_{3,4}(y)=\theta_{1,2}(W-y)$,
and $\psi^{out}=(\psi^{in})^\dag$. The Heaviside function $\theta$ reflects the vanishing of the
wavefunction at the potential walls.
The relation between amplitudes for ingoing and outgoing modes is given by
\be
\left(\begin{array}{r}
b_1\\b_2\\b_3\\b_4\end{array}\right)=s_0\left(\begin{array}{r}
a_1\\a_2\\a_3\\a_4\end{array}\right),
\ee
which defines the scattering matrix $s$, satisfying unitarity condition $s^\dag_0 s_0=I$.
Due to the geometric symmetry of the model, the scattering matrix can be
expressed as
\be
s_0=\frac{1}{2}\left(\begin{array}{rrrr}
1&0&1&0\\
0&1&0&1\\
1&0&-1&0\\
0&1&0&-1\end{array}\right)
\left(\begin{array}{rr}
s_s&0\\
0&s_a\end{array}\right)
\left(\begin{array}{rrrr}
1&0&1&0\\
0&1&0&1\\
1&0&-1&0\\
0&1&0&-1\end{array}\right).
\ee
Here $s_s$ and $s_a$ are scattering submatrices for
symmetric ($1+3$, $2+4$) and antisymmetric modes ($1-3$, $2-4$), respectively.
Since $\psi_a(y=W/2)=0$, the antisymmetric modes propagate
unperturbed along the junction, which implies $s_a=I$.

The expected form of $s_s$ is
\be
s_s=e^{i\theta}\left(\begin{array}{ll}
i\alpha&\sqrt{1-\alpha^2}\\
\sqrt{1-\alpha^2}&i\alpha\end{array}\right),
\ee
where $\alpha\in[-1,1]$ and $\theta$ is a phase value.

Our final form matrix of the $s_0$ is then
\be
\left(\begin{array}{cccc}
A&B&A&C\\
B&A&C&A\\
A&C&A&B\\
C&A&B&A
\end{array}\right)\label{mmat}
\ee
with $2A=i\alpha\exp(i\theta)$, $2B=\sqrt{1-\alpha^2}\exp(i\theta) +1$ and
$2C=\sqrt{1-\alpha^2}\exp(i\theta)-1$.
It is clear that the requirement of absence backscattering implies $\alpha=0$.

\begin{figure}
\subfigure{
\includegraphics[scale=.5]{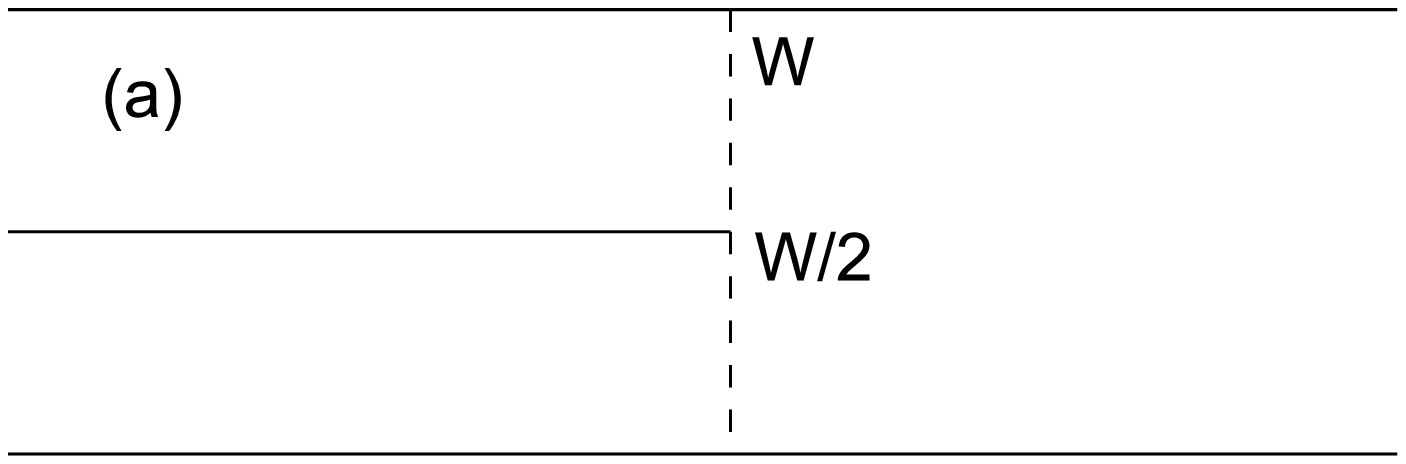} }
\subfigure{
\includegraphics[scale=.5]{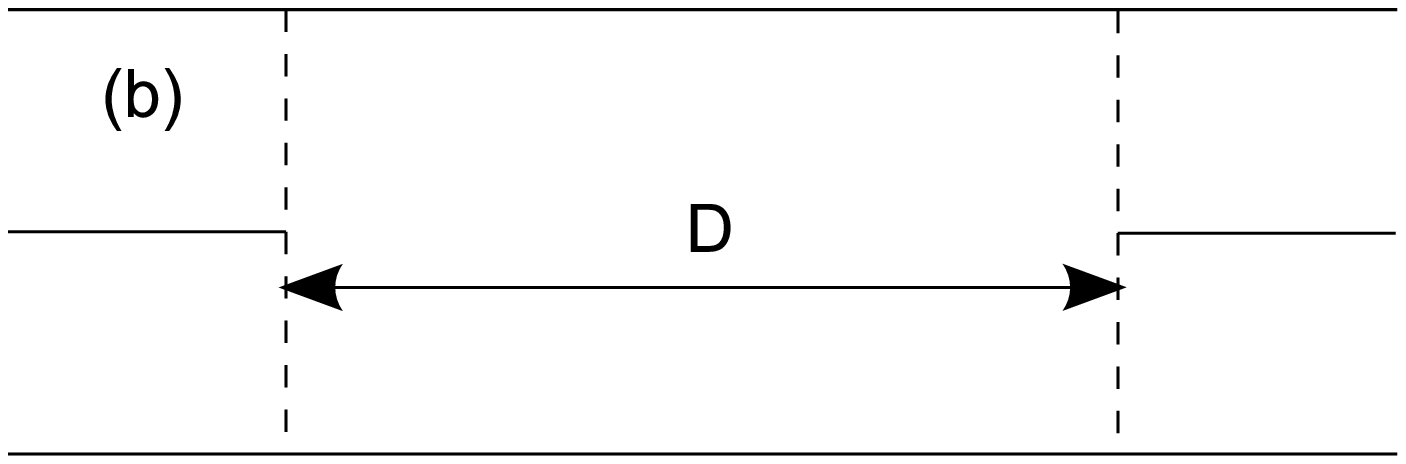}}
\caption{(a) The $Y$-wire. (b)$X$-wire made of two $Y$-wires.
Note that the symmetry allows the antisymmetric modes to propagate freely.}\label{xywi}
\end{figure}

An appropriately long $X$-junction can be seen as two $Y$-junctions depicted in Fig.~\ref{xywi}(a)
connected by
a two mode channel, as shown in Fig.~\ref{xywi}(b).
We can express the corresponding values of $\alpha$ and $\theta$ by elements of the scattering matrix
for the symmetric mode in the $Y$-junction
\be
\left(\begin{array}{ll}
a&b\\
b&c\end{array}\right),\label{may}
\ee
where
$
c=-ba^\ast/b^\ast.
$
From unitarity, we have $|a|^2+|b|^2=1$. The dependence of $a$ and $b$
on $k_FW$ can be determined numerically by properly matching propagating and evanescent modes, as
explained in Appendix C.
The parameters $\alpha$ and $\theta$ in the limit $D\gg W$ are given by
\be
\alpha=\frac{\gamma}{\sqrt{1+\gamma^2}},\:e^{i\theta}=
\frac{|b|^4e^{i(k_1-k_2)D}\sqrt{1+\gamma^2}}{(b^\ast)^2-b^2(a^\ast)^2e^{2ik_1D}},
\ee
with
\be
\gamma=2\mathrm{Im}\frac{ae^{-ik_1D}}{b^2}.
\ee

\begin{figure}
\includegraphics[scale=.6,angle=-90]{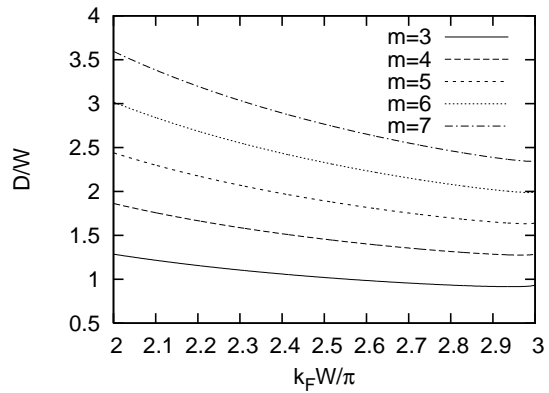}
\caption{Geometric ratios for resonances -- absence of backscattering -- given by Eq.~(\ref{dww}).}
\label{rezi}
\end{figure}

Now,
the RT resonance ($\alpha=0$) is determined by
\be
a(b^\ast)^2=b^2a^\ast e^{2ik_1D},\label{res}
\ee
and occurs at
\be
D/W=(m\pi-\mathrm{arg}(b^2/a))/k_1W\label{dww}
\ee
for $m=1,2,3\dots$.
Then
\be
\theta=(k_1-k_2)D+2\mathrm{arg} b.\:
\ee
We present a few lines of resonances for $k_FW/\pi\in[2,3]$
in Fig.~\ref{rezi}.  We stress that actual lines differ a little from Eq.~(\ref{dww}) due to the approximation $D\gg W$.
In general, in order to determine the exact positions of resonances, a residual contribution of evanescent modes in the middle part has to be taken into account.

\begin{figure}
\includegraphics[scale=.5]{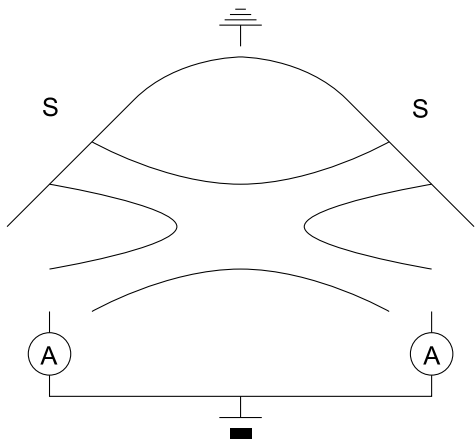}
\caption{The geometry of an $X$ junction connected to a coherent superconductor (S).}\label{x_junction}
\end{figure}

\begin{figure}
\includegraphics[scale=.6,angle=-90]{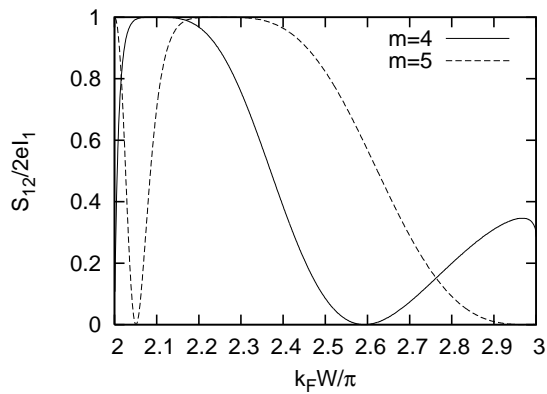}
\caption{The cross shot noise along the resonances given in Fig.\ref{rezi}}\label{rezin}
\end{figure}

The junction is connected to the superconductor as shown on Fig.~\ref{x_junction}.
The predicted magnitudes of the cross shot noise (\ref{noi}) along the RT resonances given by (\ref{dww}) are presented in Fig.~\ref{rezin}. Note that the noise magnitude reaches the maximum values
given in Eq.~(\ref{subpf}).

\begin{figure}
\includegraphics[scale=.25]{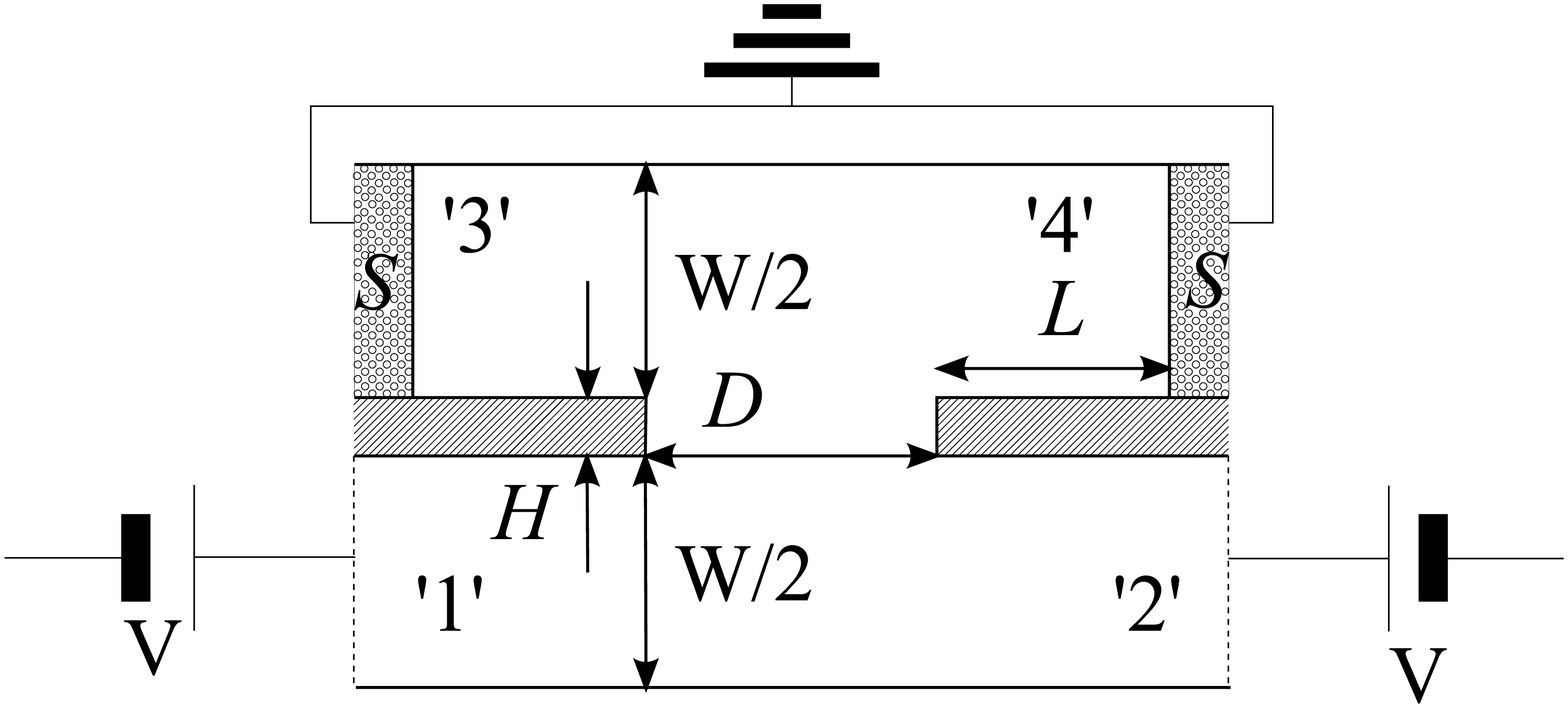}
\caption{The realistic junction with superconductor}\label{xwz}
\end{figure}

\section{Further modifications}

The maximal cross shot noise shown in Fig.~\ref{rezin}
requires not only ideal geometry and
transparency but also tuning both $k_FW$ and $D/W$
according to (\ref{dww}). Moreover, non-zero temperature also usually decreases cross correlations.

Accordingly, we have considered the scattering problem for a more realistic device shown in Fig.~\ref{xwz}, assuming
an imperfect interface, taken into account within the Blonder, Tinkham and Klapwijk (BTK)\cite{btk} model, varying
the distance $L$ between the superconductor and the junction and/or allowing for a non-zero width $H$ of the internal gate between the terminals.

The scattering problem is solved by mode matching, taking into account not completely vanishing evanescent modes in the middle region.
A single contact supports one mode for $k_FW\in]2\pi,4\pi[$.
The transparency of non-ideal superconductor-normal metal interface in the BTK
model reads $\Gamma=1/(1+Z^2)$.

\begin{figure}
\includegraphics[scale=.7,angle=-90]{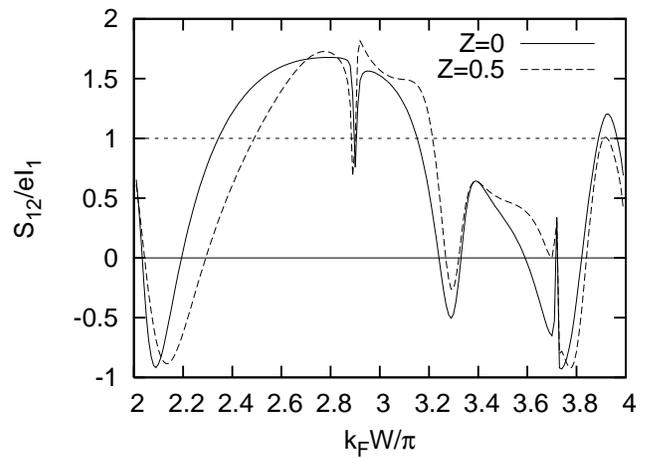}
\caption{The cross shot noise for ideal and non-ideal superconducting-normal interface for $D=L=W=10H$
as described in Fig.~\ref{xwz} for $Z=0$ and $0.5$. Note that in both cases $S_{12}$ can exceed
maximal value $e\bar{I}_1$ for a single-mode superconductor.}\label{dl2}
\end{figure}

We present the results in Fig.~\ref{dl2} for $H=W/10$, $D=L=W$,
$Z=0$, and $Z=0.5$.
The resonance seem to be wider and more robust
for $k_FW<3\pi$ but large $L$ can turn it into several
local maxima. However, $S_{12}$ still exceed $e\bar{I}_1$
for some range of parameters in contrast to previous results.\cite{tor00}

It is clear that to get strong positive cross shot noise one should be able to tune the parameters of the junction since the
noise is highly sensitive to changes. However, the general
tendency is that the narrower the junction is, the more stable the noise magnitude is.

\begin{figure}
\includegraphics[scale=.5]{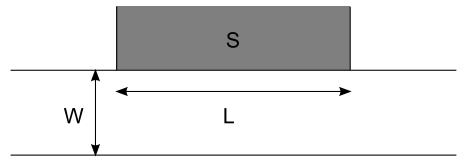}
\caption{The $T$-junction. Superconductor is attached from above to a normal quantum point contact }\label{tju}
\end{figure}

\begin{figure}
\includegraphics[scale=.7,angle=-90]{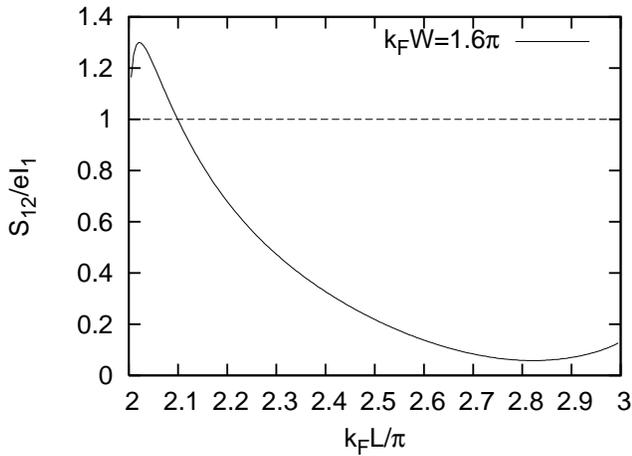}
\caption{The cross shot noise of $T$-junction for some range of parameters.
Note that the curve exceeds $1$ -- the maximal value for all single modes}\label{tres}
\end{figure}

The presented $X$-junction may be difficult to realize
experimentally. Therefore, we have analyzed also the case
of $T$-junction with two modes going into the superconductor
and single modes in normal terminals (Fig.~\ref{tju}).
As shown in Fig.~\ref{tres}, the cross noise cannot reach
the maximum $2e\bar{I}_1$ in such a geometry but still, for a certain range of parameters, it is larger than $e\bar{I}_1$.
This large magnitude cannot be attained for any three-single-mode junction,\cite{tor00} in 
chaotic cavity\cite{Sam00} or semiclassical regime.
\cite{bro}


\section{Summary}

We have proposed examples of X-junctions that
exhibit, according to our theoretical results, large magnitudes of positive cross shot noise.
Such a large magnitude could not be attained in the previously studied cases, such as three terminal devices with single modes
in each leg or chaotic cavities containing many modes. The presented examples require
separate connections to the phase coherent superconductor.
One can, however, consider another example -- simple Y or T junction, in which the leg connected to the
superconductor contains at
least two modes. The cross noise in this case can be also positive but not so large as
in the X-junction.  Nevertheless, in principle
 one should always be able to modify every four-mode junction
in order to get maximal noise - ideal splitting of electrons.  Hence,
narrow wires are promising when searching for
considerable positive cross correlations.
Lastly, we would like to mention
that some experiments to measure the cross shot noise in
junctions discussed here are in preparation.

\section*{Acknowledgments}
We acknowledge support by the Eurocores/ESF grant
SPINTRA (ERAS-CT-2003-980409) and the German
Research Foundation (DFG) through SFB 513, SFB 767,
and SP 1285 Semiconductor Spintronics. We are grateful to W. Belzig for helpful remarks.

\section*{Appendix A}
\renewcommand{\theequation}{A.\arabic{equation}}
\setcounter{equation}{0}

The long time properties of electronic transport are well described by full counting statistics,
\cite{les,naz} with a generalization to the normal metal-superconductor interface. \cite{muz}
We consider the particle transfer statistics through a mesoscopic junction at given temperature $T$ and
voltage bias $V$, without interactions. The junction has $m$ terminal/modes and a particle detector can be placed
at each of them. During the measurement process a detector at the terminal/mode $\alpha$ registers the difference
between numbers of particles outgoing from and ingoing to the junction $N_\alpha$.
Here $\alpha$ denotes terminal, mode, spin and particle type (electron or hole).
A set of registered numbers
 $\boldsymbol N=(N_1,\dots,N_m)$ occurs with a probability $p(\boldsymbol N)$. Instead of probability, a very convenient tool
 to describe statistical properties of a probability distribution is the generating function.
\be
e^{S(\boldsymbol\lambda)}=\langle e^{i\boldsymbol\lambda\cdot\boldsymbol N}\rangle=
\sum_{\boldsymbol N} p(\boldsymbol N)e^{i\boldsymbol\lambda\cdot\boldsymbol N}.
\ee
Here $\boldsymbol\lambda=(\lambda_1,\dots,\lambda_m)$ is the vector of counting variables $\lambda_\alpha$.
Using this form, it is straightforward to express averages and correlation functions as
\be
\langle N_\alpha\rangle=-i\frac{\partial S}{\partial\lambda_\alpha},
\:\langle\delta N_\alpha\delta N_\beta\rangle=-\frac{\partial^2 S}{\partial\lambda_\alpha\partial\lambda_\beta}
\label{full}
\ee
for $\delta N_\alpha=N_\alpha-\langle N_\alpha\rangle$.
On the other hand, the generating function at long times, $t_0\gg h/(|eV|+k_BT)$ is given by Levitov-Lesovik formula\cite{les}
\be
S=\frac{t_0}{h}\int dE\;\ln\mathrm{det}(1+n(s^\dag e^{i\Lambda}se^{-i\Lambda}-1)).
\ee
Here $\Lambda$, $n$, $s$ and $s^\dag$  are $m\times m$ matrices.
The first two are diagonal with $\Lambda_{\alpha\alpha}=\lambda_\alpha$ and $n_{\alpha\alpha}=n_\alpha$. The latter describes occupation numbers, $n_\alpha=(1+\exp((E-q_\alpha V_\alpha)/k_BT))^{-1}$
with $q_\alpha=-e$ for electrons and $+e$ for holes and $V_\alpha$ is the bias voltage.
Exploiting the identity $\tr\ln=\ln\det$ the averages and correlations can be written as
\ba
&&
\langle N_\alpha\rangle=\frac{t_0}{h}\int dE\;\tr P_\alpha(sns^\dag-n),\label{nne}\\
&&
\langle \delta N_\alpha\delta N_\beta\rangle=\frac{t_0}{h}\int dE\;\tr[sns^\dag P_\alpha(1-sns^\dag)P_\beta\nonumber\\
&&
+n(1-n)(P_\alpha P_\beta-s^\dag P_\alpha s P_\beta-s^\dag P_\beta sP_\alpha)]=\nonumber\\
&&\frac{t_0}{h}\int dE\;\tr[n(s^\dag P_\alpha s-P_\alpha)(1-n)(s^\dag P_\beta s- P_\beta)]\nonumber
\ea
where $P_\alpha$ denotes the projection on the mode $\alpha$ so that $P_\alpha$ commutes with $n$.

In this limit transport can be interpreted in terms of a series of elementary transport events. At time
period $h/\Delta E$, detectors at each terminal can register an incoming particle, $\Delta N_\alpha=-1$,
and/or an outgoing particle, $\Delta N_\alpha=+1$, or nothing.
The probability value of the event that the set $I$ of ingoing particles is converted to the set $O$ of outgoing particles
is given by
\be
p(I\to O)=|\mathrm{det}\left(P_{I}sP_{O}\right)|^2\prod_{\alpha\in I}n_\alpha\prod_{\alpha\notin I}(1-n_\alpha),
\label{prob}
\ee
where $P_{I}$ and $P_{O}$ denote the projections on the given set of modes/terminals.
As the long time statistics can be interpreted classically, it also satisfies CBS inequality
\be
\langle\delta Q_A\delta Q_B\rangle^2\leq\langle(\delta Q_A)^2\rangle\langle(\delta Q_B)^2\rangle
\label{Schwarzmaster}
\ee
for $\delta Q_X=\sum_{\alpha\in X}q_\alpha\delta N_\alpha$ and arbitrary sets $A$ and $B$.
To prove the inequality (\ref{Schwarzmaster}) it is enough to show that
$
\langle(\delta Q_A-z\delta Q_B)^2\rangle\geq 0
$
for $z=\langle\delta Q_A\delta Q_B\rangle/\langle(\delta Q_B)^2\rangle$. Let us define
$P_z=\sum_{\alpha\in A}q_\alpha P_\alpha-z\sum_{\alpha\in B}q_\alpha P_\alpha$. Then
\ba
&&\langle(\delta Q_A-z\delta Q_B)^2\rangle=\nonumber\\
&&\frac{t_0}{h}\int dE\;\tr[n(s^\dag P_z s-P_z)(1-n)(s^\dag P_z s- P_z)]\nonumber\\
&&=\frac{t_0}{h}\int dE\;\tr K^\dag K\label{kkk}
\ea
for $K=\sqrt{n}(s^\dag P_z s-P_z)\sqrt{1-n}$ since $n^\dag=n$. The trace of a Hermitian square is always positive, which completes the proof.
Moreover, at zero temperature in the normal case, the noise is always sub-Poissonian if every terminal is either grounded or at the same voltage $V$,
\be
\langle (\delta Q_V)^2\rangle\leq |e\langle Q_V\rangle|\label{subp}
\ee
for $Q_V=\sum_{\alpha\in V}q_\alpha N_\alpha$ with summation over all terminals at $V$.
To prove it, we use the fact that $n=\theta(-E)$ or $n=\theta(eV-E)$ for
terminals at $0$ or $V$, respectively.
Then, using (\ref{nne}) we get
\ba
&&
\langle Q_V\rangle=\frac{t_0}{h}\int_0^{eV} edE\;\tr ss^\dag,\label{nne1}\\
&&
\langle (\delta Q_V)^2\rangle=\frac{t_0}{h}\int_0^{eV} e^2dE\;\tr[ss^\dag-(ss^\dag)^2]\nonumber
\ea
We get (\ref{subp}) since $\tr(ss^\dag)^2\geq 0$.
In the case of the ground replaced by the superconductor, the  problem reduces
to the normal case if NS surface is treated as a mirror, doubling the number of terminals for different quasiparticles. The total noise is
\be
\langle (\delta Q)^2\rangle=\langle (\delta Q_+)^2\rangle
+\langle (\delta Q_-)^2\rangle+2\langle\delta Q_+\delta Q_-\rangle
\ee
where $+$ and $-$ denote real and mirrored terminals, respectively.
From CBS inequality (\ref{Schwarzmaster}) we get 
\be
\langle (\delta Q)^2\rangle\leq 2(\langle (\delta Q_+)^2\rangle
+\langle (\delta Q_-)^2\rangle)
\ee
Using (\ref{subp}) we get
\be
\langle (\delta Q)^2\rangle\leq 2(|e\langle Q_+\rangle|+|e\langle Q_-\rangle|)=
2|e\langle Q\rangle|
\ee
because $\langle Q_+\rangle=-\langle Q_-\rangle$.

\section*{Appendix B}
\renewcommand{\theequation}{B.\arabic{equation}}
\setcounter{equation}{0}

The CBS inequality (\ref{Schwarzmaster}) becomes equality only when $K=0$ in
(\ref{kkk}). For a finite temperature the ideal splitting gives the condition
$s^\dag P_{z=1} s=P_{z=1}$.  The only possible scattering matrix in the basis $(1e,2e,1h,2h)$ has the block structure
\be
s=\left(\begin{array}{rrrr}
\ast &0 &0 &\ast\\
0&\ast &\ast &0\\
0&\ast &\ast &0\\
\ast &0 &0 &\ast\end{array}\right)
\ee
This means that the trace of $s_{eh}$  must vanish.
However, from (\ref{supm}), it implies $\tr(2-T)T=0$ for
$T=tt^\dag=1-rr^\dag$. For the fact that eigenvalues of $T$ lie
 between $0$ and $1$, the only possibility would be $t=0$ but this excludes
 superconductor completely and gives zero noise.
 
 At zero temperature and finite voltage the condition $K=0$
 has other solutions because $n_e\neq n_h$. The new requirement is
 \be
 (s^\dag P_{z=1} s)_{eh}=0\label{cond}
 \ee
 In our special case (\ref{rmm}), using (\ref{supm}) we get the condition.
 \be
A^\ast(B^2-A^2)=A.\label{aab}
\ee
The unitarity of $s_0$ imposes conditions
$|A+B|\leq 1$ and $|A-B|\leq 1$. Together with (\ref{aab}) it implies either $A=0$ or $|A+B|=1=|A-B|$. The latter possibility again gives $t=0$ so we are left with the former one.

In the case of three mode junction we have
\be
t=\left(\begin{array}{r}
t_1\\
t_2\end{array}\right)
\ee
 The unitarity condition yields $tt^\dag$ proportional to identity matrix if $A=0$ in $r$.
Hence, $|t_1|^2=|t_2|^2$ and $t_1^\ast t_2=0$ so $t_1=t_2=0$ and the superconductor decouples.

\section*{Appendix C}
\renewcommand{\theequation}{C.\arabic{equation}}
\setcounter{equation}{0}

The symmetric wavefunction can be
reduced to the interval  $y\in[0,W/2]$
since
\be
\psi(x,y)=\psi(x,W-y)
\ee
For $y\in[0,W/2]$ we have the following decomposition
into $N+1\to\infty$ evanescent modes
\ba
&\psi_L=&\sin(2\pi y/W)(Ae^{ik_2x}+Be^{-ik_2x})+\nonumber\\
&&
\sum_{j=2}^{N+1}(-1)^j\sin(2j\pi y/W)E_{2j}e^{\kappa_{2j}x}\nonumber\\
&
\psi_R=&\sin(\pi y/W)Ce^{ik_1x}+\\
&&\sum_{l=1}^N(-1)^l\sin((2l+1)\pi y/W)
 E_{2l+1}e^{-\kappa_{2l+1}x}\nonumber
\ea
where
\be
\kappa_nW=\sqrt{n^2\pi^2-(k_FW)^2},\:n=3,4,\dots
\ee
The boundary conditions (integration with $\chi_{2j}(y)$, $y\in[0,W/2]$) are
\ba
&&\int_0^{W/2}dy\;\sin\left(\frac{2j\pi y}{W}\right)(\psi_L(0,y)-\psi_R(0,y))=0\nonumber\\
&&
\int_0^{W/2}dy\:\sin\left(\frac{2j\pi y}{W}\right)\partial_x(\psi_L(0,y)-\psi_R(0,y))=0\nonumber
\ea
We make use of trigonometric identities 
and integrals
\ba
&&\int_0^{W/2}d y\:\sin^2(2j\pi y/W)=W/4
\\
&&
\int_0^{W/2}d y\:\sin(2j\pi y/W)\sin((2l+1)\pi y/W)=\nonumber\\
&&\frac{W}{2\pi}(-1)^{l+j}
\frac{4j}{(2l+1)^2-4j^2}\nonumber
\ea
We get the following equations for $A=1$, $B$, $C$, $E_3$, $E_4$,...$E_{2N+2}$
and $j=2,\dots N+1$,
\ba
&&A+B=\frac{8}{3\pi}C-\sum_{l=1}^N\frac{8}{\pi}
\frac{E_{2l+1}}{(2l+1)^2-4}
\nonumber\\
&&
E_{2j}=\frac{8j}{\pi}\frac{C}{1-4j^2}+\sum_{l=1}^N
\frac{8j}{\pi}\frac{E_{2l+1}}{(2l+1)^2-4j^2},
\nonumber\\
&&
ik_2(A-B)=\frac{8}{3\pi}ik_1C+\sum_{l=1}^N\frac{8}{\pi}
\frac{\kappa_{2l+1}E_{2l+1}}{(2l+1)^2-4}
\\
&&
\kappa_{2j}E_{2j}=\frac{8j}{\pi}\frac{ik_1C}{1-4j^2}-\sum_{l=1}^N
\frac{8j}{\pi}\frac{\kappa_{2l+1}E_{2l+1}}{(2l+1)^2-4j^2},
\nonumber
\ea
The elements of the matrix (\ref{may}) are given by
\be
a=\frac{B}{A},\:b=\sqrt{\frac{k_1}{k_2}}\frac{C}{A}.
\ee
Note that for finite $N$ the unitarity condition $|a|^2+|b|^2=1$ may be not exactly satisfied.

\end{document}